\documentclass[conference]{IEEEtran}
\IEEEoverridecommandlockouts
\usepackage{amsmath,amsfonts}
\usepackage{algorithmic}
\usepackage{graphicx}
\usepackage{textcomp}
\usepackage{xcolor}
\usepackage{xspace}
\usepackage{pifont}
\usepackage{hyperref}
\usepackage{xspace}
\usepackage{listings}
\usepackage{subcaption}

\newcommand*{\ie}{i.e.,\@\xspace}
\newcommand*{\eg}{e.g.,\@\xspace}

\newcommand*{\etal}{et. al\@\xspace}

\newcommand*{\tool}{CodeXHug\@\xspace}

\newcommand*{\HGF}{Hugging Face\@\xspace}
\newcommand*{\GH}{GitHub\@\xspace}
\newcommand*{\RM}{README\@\xspace}

\newcommand*{\numTot}{681,682\@\xspace}
\newcommand*{\numNonull}{262,670\@\xspace}
\newcommand*{\numFilter}{20,545\@\xspace}
\newcommand*{\numTagged}{17,760\@\xspace}
\newcommand*{\numWithCode}{7,325\@\xspace}
\newcommand*{\numFiles}{372,063\@\xspace}

\lstdefinelanguage{json}{
    basicstyle=\scriptsize\ttfamily, 
    numbers=left, 
    numberstyle=\tiny\color{gray}, 
    stepnumber=1, 
    numbersep=5pt, 
    showstringspaces=false, 
    breaklines=true, 
	captionpos=b,
    frame=lines, 
    backgroundcolor=\color{lightgray!20}, 
    literate=
     *{0}{{{\color{blue}0}}}{1}%
      {1}{{{\color{blue}1}}}{1}%
      {2}{{{\color{blue}2}}}{1}%
      {3}{{{\color{blue}3}}}{1}%
      {4}{{{\color{blue}4}}}{1}%
      {5}{{{\color{blue}5}}}{1}%
      {6}{{{\color{blue}6}}}{1}%
      {7}{{{\color{blue}7}}}{1}%
      {8}{{{\color{blue}8}}}{1}%
      {9}{{{\color{blue}9}}}{1}%
      {:}{{{\color{red}:}}}{1}%
      {,}{{{\color{red},}}}{1}%
      {"}{{{\color{brown}"}}}{1}%
      {true}{{{\color{violet}true}}}{1}%
      {false}{{{\color{green}false}}}{1}%
      {null}{{{\color{magenta}null}}}{1}%
}
\def\BibTeX{{\rm B\kern-.05em{\sc i\kern-.025em b}\kern-.08em
    T\kern-.1667em\lower.7ex\hbox{E}\kern-.125emX}}
\begin{document}

\title{Generate with CodeXHug: A Dataset to Enhance Model Cards with Code Usage Patterns \\}

\author{\IEEEauthorblockN{Stefano Palombo}
\IEEEauthorblockA{
\textit{University of l'Aquila, Italy }\\
stefano.palombo@student.unviaq.it}
\and
\IEEEauthorblockN{Claudio Di Sipio}
\IEEEauthorblockA{
\textit{University of l'Aquila, Italy}\\
claudio.disipio@univaq.it}
\and
\IEEEauthorblockN{Juri Di Rocco}
\IEEEauthorblockA{
\textit{University of l'Aquila, Italy }
\\ juri.dirocco@univaq.it}

\and
\IEEEauthorblockN{Davide Di Ruscio}
\IEEEauthorblockA{
\textit{University of l'Aquila, Italy}\\
davide.diruscio@univaq.it}
}

\maketitle

\begin{abstract}
    Pre-trained models (PTMs) are becoming increasingly popular in the software engineering community. Their usage is facilitated by model repositories, \eg \HGF, which collect, store, and maintain a wide range of PTMs. However, the actual adoption of these models in real-world projects is still an open question, \ie many of them are used in toy projects or simply as a mirror for the HF repository. In addition, most of the available model cards and textual documents that contain critical information about their usage do not include explanatory code patterns, thus increasing the difficulty for newcomers.  Thus, we see the need for a curated codebase related to PTMs to support developers and practitioners who are interested in using them in their projects.
    In this paper, we present \tool, a curated dataset of \HGF PTMs exploited in the \GH ecosystem and the related code usage patterns. Starting from the latest HF dump, we first conduct a data curation to collect PTMs with a tag and a model card. Then, the \GH platform has been queried to find actual usages of the identified PTMs, resulting in \numWithCode different models and \numFiles Python files. 
    To demonstrate a concrete application of \tool, we propose a usage scenario focused on extracting representative code usage patterns for specific PTMs through a statistical analysis and clustering techniques applied to relevant code snippets.
    
   
    Finally, we discuss the research opportunities enabled by \tool and the implications of our findings for the software engineering community. In particular, we believe that \tool can be used to support different tasks related to the usage of PTMs, from providing developers with concrete usage examples to generating curated model cards, contributing to the development of next-generation solutions based on top of PTMs
\end{abstract}


\section{Introduction}
With the advent of foundation models, \eg large language models (LLMs) \cite{10.1145/3695988} or pre-trained models (PTMs) \cite{HAN2021225}, the field of software engineering (SE) has been significantly transformed, fostering the automation of several tasks leveraging those cutting-edge technologies. PTMs can be seen as specialized off-the-shelf components that support specific tasks, outperforming traditional techniques in many cases \cite{tufano_using_2022,ding_can_2022,zhang_using_2022}. In addition, the proliferation of open \textit{model repositories} has facilitated their usage in practice. \HGF is the largest model repository for SE tasks, providing a wide range of PTMs to support more than thirty different tasks, \eg text generation, image classification, or code summarization. Each model is labeled with \textit{pipeline tags}, thus facilitating the repository's browsing and increasing the models' discoverability given the current software engineering task. 
In addition, \HGF also provides dedicated capabilities that allow developers to upload their models, categorize them, and upload relevant information and meta-data using the \textit{model card} \cite{10.1145/3287560.3287596}. This README-like document provides detailed instructions on how to run and install store models. Nevertheless, recent research has shown that different information is missing, \eg carbon emission \cite{castano_analyzing_2023} or license information \cite{10.1145/3643916.3644412}, which may hinder the adoption of PTMs in practice. In particular, several model cards lacks of concrete \textit{code usage examples}, thus making the integration in existing projects challeging especially for non-expert users. In this respect, open-source software (OSS) ecosystems, such as \GH, can come in handy, providing a valuable codebase for thos cutting edge models. Neverthelss, indetifying useful code snippets for PTMs is still an open issue, as many of them are used in toy projects or simply as a mirror for the HF repository.

To fill this gap, we propose \tool, a curated dataset of PTMs stored in HF that have been used in \GH repositories. To collect data, we leverage the latest version of the \HGF dump provided by the HF community project \cite{ait_hfcommunity_2023}. We first filter out unpopular models using the number of downloads as a proxy, and then we identify the projects that exploit the elicited PTMs. To this end, we rely on a well-adopted Python library, \ie PyGithub \cite{pygithub}, to collect relevant data from the identified PTMs. We ended up with \numWithCode different models and \numFiles Python files. In addition, we applied a set of qualtiative filters to enable the generation of code usage patterns by combining KNN clustering algorithm and Llama 3 model. 


To the best of our knowledge, \tool is the first dataset that provides a comprehensive overview of the actual usage of PTMs in the \GH ecosystem, thus enabling different research opportunities, \eg code assistants for developing  PTMs, PTMs usage analysis, and model card generation.

The contributions of this paper are as follows:

\begin{itemize}
    \item A curated dataset, named \tool, that maps existing code snippets to most popular PTMs available on HF platform, available on the Zenodo open-access research data repository~\cite{di_sipio_2024_14267550}.
    \item A qualitative analysis of the collected code snippets to provide developers with concrete usage examples for PTMs.
    \item An explanatory example of how to use \tool to predict the usage of PTMs in real-world projects leveraging LLM and clustering techniques.
\end{itemize}

The paper is structured as follows. Section \ref{sec:background} presents an explanatory example to motivate the work. Section \ref{sec:curation} describes the data collection process while Section \ref{sec:analysis} provides an overview of the dataset, including a statistical analysis of the collected data. We present a concrete usage of \tool for generating code usage patterns in Section \ref{sec:usage}. Section \ref{sec:threats} discusses the threats to validity of our study. Finally, Section \ref{sec:related} presents related works and Section \ref{sec:conclusion} concludes the paper.

\label{sec:introduction}

\section{Background and Motivation}
While Hugging face offers different capabilties to store, maintain, and document PTMs, recent studies highlight the limitation in terms of relevant information for developers, \eg discrepancies in the documentations \cite{montes_discrepancies_2022}, or lack of user liceses \cite{10.1145/3643916.3644412}. 

\begin{figure}[h]
    \centering
    \begin{minipage}{0.45\textwidth}
        \centering
        \includegraphics[width=\textwidth]{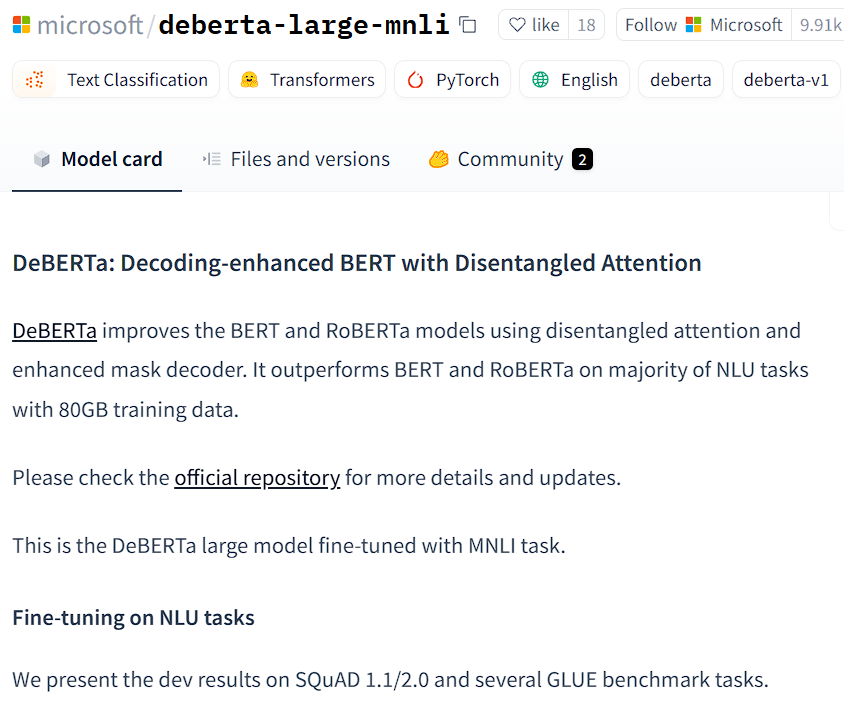}
        \subcaption{Deperta model card}
        \label{fig:missing_info_no_code}
    \end{minipage}
    \hfill
    \begin{minipage}{0.45\textwidth}
        \centering
        \includegraphics[width=\textwidth]{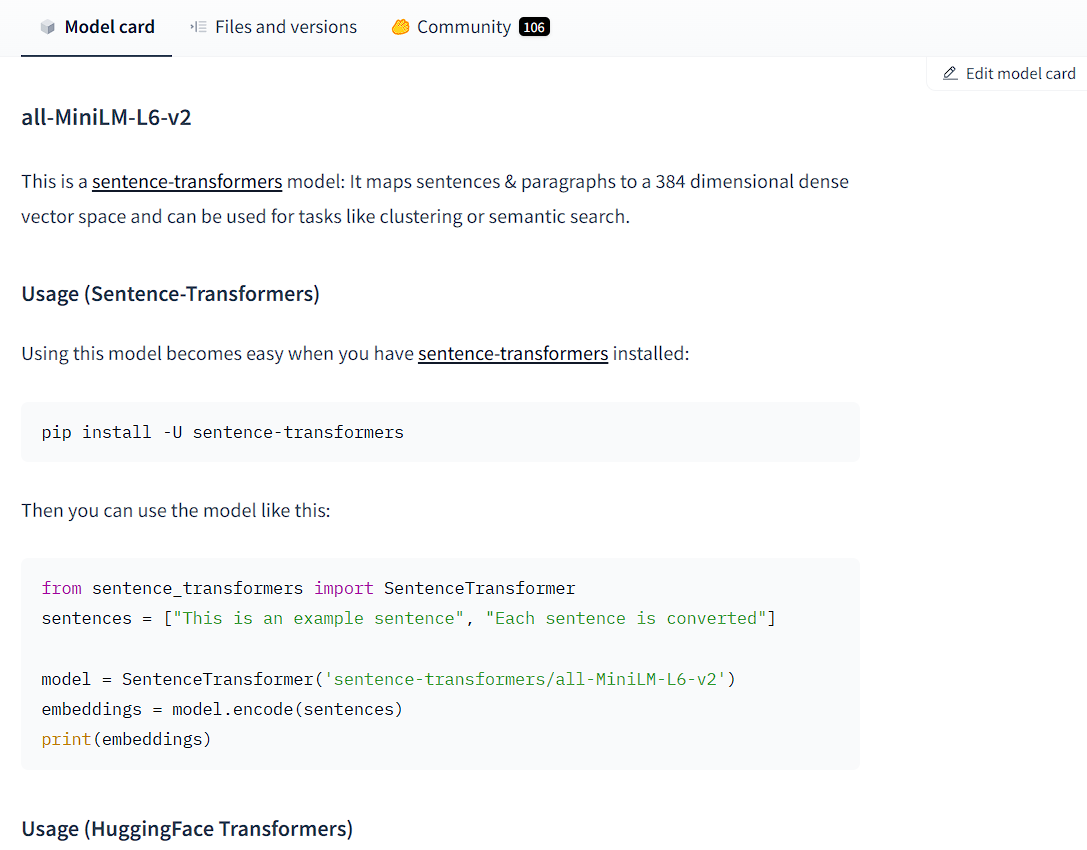}
        \subcaption{all-MiniLM-L6-v2 model card}
        \label{fig:missing_info_with_code}
    \end{minipage}
    \caption{Model cards with and without code usage examples} 
    \label{fig:missing_info} 
\end{figure}

In particular, the lack of code usage examples in model cards can make difficult their reusage in practice, especially for non-expert developers. Figure \ref{fig:missing_info_no_code} shows the model card of the \texttt{deberta-largew-mnli} PTM provided by Microsoft. It is worth mentioning that the document do not contains any usage examples even though the model is very popular according to HF statistics. In contrast, Figure \ref{fig:missing_info_with_code} shows a model card related to the \texttt{all-MiniLM-L6-v2} with code usage examples. In particular, this model card is enriched with useful scripts for installing and a basic usage of the model, thus making it easier for developers to exploit it in their projects.

\label{sec:background}

\section{Data collection} 
\begin{figure}[b!]
    \centering
    \includegraphics[width=0.5\textwidth]{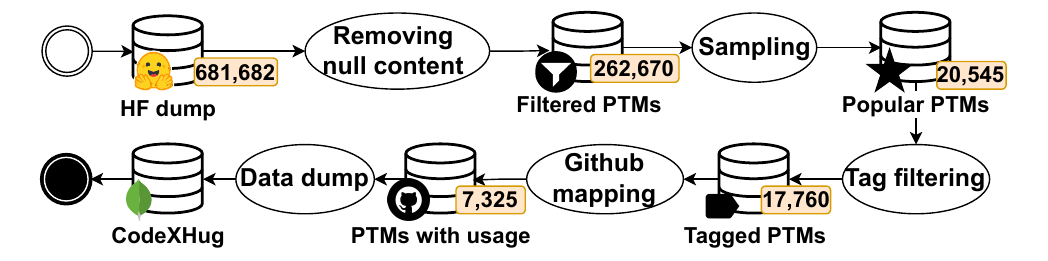}
    \caption{The \tool collection process}
    \label{fig:tool}
\end{figure}

Figure \ref{fig:tool} deipicts the \tool collection process. The process begins with a data cleaning step, where PTMs with null content, \ie those lacking tags or model cards, are filtered out. Next, we focus on selecting a sample of the most popular PTMs based on download counts. To ensure a balanced dataset, we identify the 13 most representative categories and then proceed to search for code usage in \GH. 

\subsection{\tool data model}

\begin{figure}[t!]
    \centering
    \includegraphics[width=\linewidth]{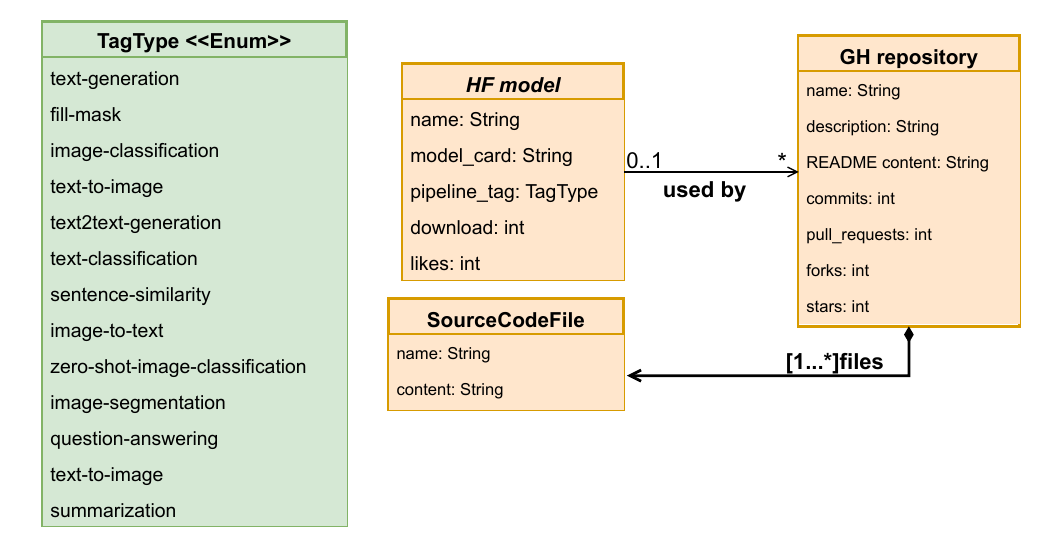}
    \caption{The \tool data model}
    \label{fig:model}
\end{figure}

Figure \ref{fig:model} shows the \tool data model consisting of two main entities, \ie \texttt{HF model} and \texttt{GH repository}. The former contains the name of the PTM, the pipeline tag, the model card, and the number of likes and downloads. The latter contains the name of the repository, the topics, the description, the \RM content, and the number of commits, pull requests, forks, and stars. In addition, we store the files of each repository, \ie the file name and the URL. In addition, we report the 13 different categories that are included in \tool dataset as an enumeration since they refer to the pipeline tags available on HF which is a closed set of values.

\subsection{Data cleaning and sampling} 

The first step involves the data cleaning and curation of the HF dump provided by the HF community project \cite{ait_hfcommunity_2023}, namely the one released on June 2024.
To interact with the dataset, we rely on Python MySQL connector \cite{mysql_connector_python} to query the database and retrieve the relevant information.
The dump contains \numTot different models even though several missing entries are present, \ie null values for tags or model cards. Since those data are relevant for our analysis, we filter out the PTMs with null content, resulting in \numNonull PTMs and the corresponding metadata, \ie pipeline\_tag, model\_card, and number of likes, and download. Afterward, we select a sample of 10\% the most popular PTMs in terms of download, \ie \numFilter PTMs. The rationale behind this choice is that the most downloaded PTMs are more likely to be used in real-world projects, thus providing a more representative sample for our analysis.

\subsection{Tag filtering} Afterward, we investigated to what categories the collected PTMs belongs. In particular, we identify 13 different categories such that we include \textit{i)} the most popular ones in terms of the number of downloads and \textit{ii)} less popular ones to guarantee a diverse enough dataset. 

It is worth noting that we check the number of PTMs for each category with the aim of balancing the dataset as much as possible. This additional filtering step ends up with \numTagged PTMs.

\subsection{\GH mapping}

This step involves the search on the whole \GH platform, even though we introduce quality filters in the query to retrieve only relevant repositories for our analysis. To this end, we rely on the PyGithub library \cite{pygithub} to interact with the \GH API and collect the relevant data. 

First, we search the name of each PTM and limit our search to repositories written in Python, as the examined PTMs are mainly imported and tested using the HF utilities\footnote{\url{https://huggingface.co/docs/transformers/autoclass_tutorial}}. Second, we queried the \RM file content using the \textit{pipeline\_tag} element to further remove possible false positives. After the relevant PTMs have been identified, we retrieve the source code files, \ie the ones ending with \textit{.py}, by limiting the search to 1,000 for each project to avoid that big repository could lead to an unbalanced number of samples in the dataset. In addition, we collect also the \RM content of the project since it may contain additional usage information. In the scope of this paper, we store only the URLs for each file even though we provide a dedicated Pythoh function in the supporting online appendix that retrieves the whole content. Listing \ref{lst:data} shows the explanatory structure of the \textit{sentence-transformers/all-MiniLM-L12-v2} PTM used by \textit{run-llama/llama\_index} GH project. In addition to files and \RM, we collect relevant metadata, \ie number of commits, pull requests, forks, and stars, to provide a comprehensive overview of the project. This mapping phase ends up with \numWithCode different models and \numFiles Python files contained into 71,748 repositories. 

\begin{lstlisting}[caption=Example of retrieved data,language=json, label=lst:data]
"sentence-transformers/all-MiniLM-L12-v2": {
"run-llama/llama_index": {
    "topics": <list of GH topics>,
    "description": "LlamaIndex is a data framework for your LLM applications",
    "readme": <content>",
    "numbers of commits:": 5494,
    "number of pull requests": 6146,
    "number of forks": 5195,
    "number of stars": 36388,
    "files": [
        {
            "file_name": "bench_embeddings.py",
            "file_url": "<file_url>"
        }
    ]
}
    \end{lstlisting}


To further facilitate the analysis and the reproducibility of our study, we store the collected data in a MongoDB database \cite{mongodb} as it provides faster access compared to SQL databases. 
The provided dump is composed of two collections, \ie \textit{models} and \textit{files}, where the former contains the name of PTM and pipeline\_tag while the latter contains the content of each file encoded in UTF-8 standard.

\label{sec:curation}

\section{Data overview}

\begin{figure}[h!]
    \centering
    \includegraphics[width=\linewidth]{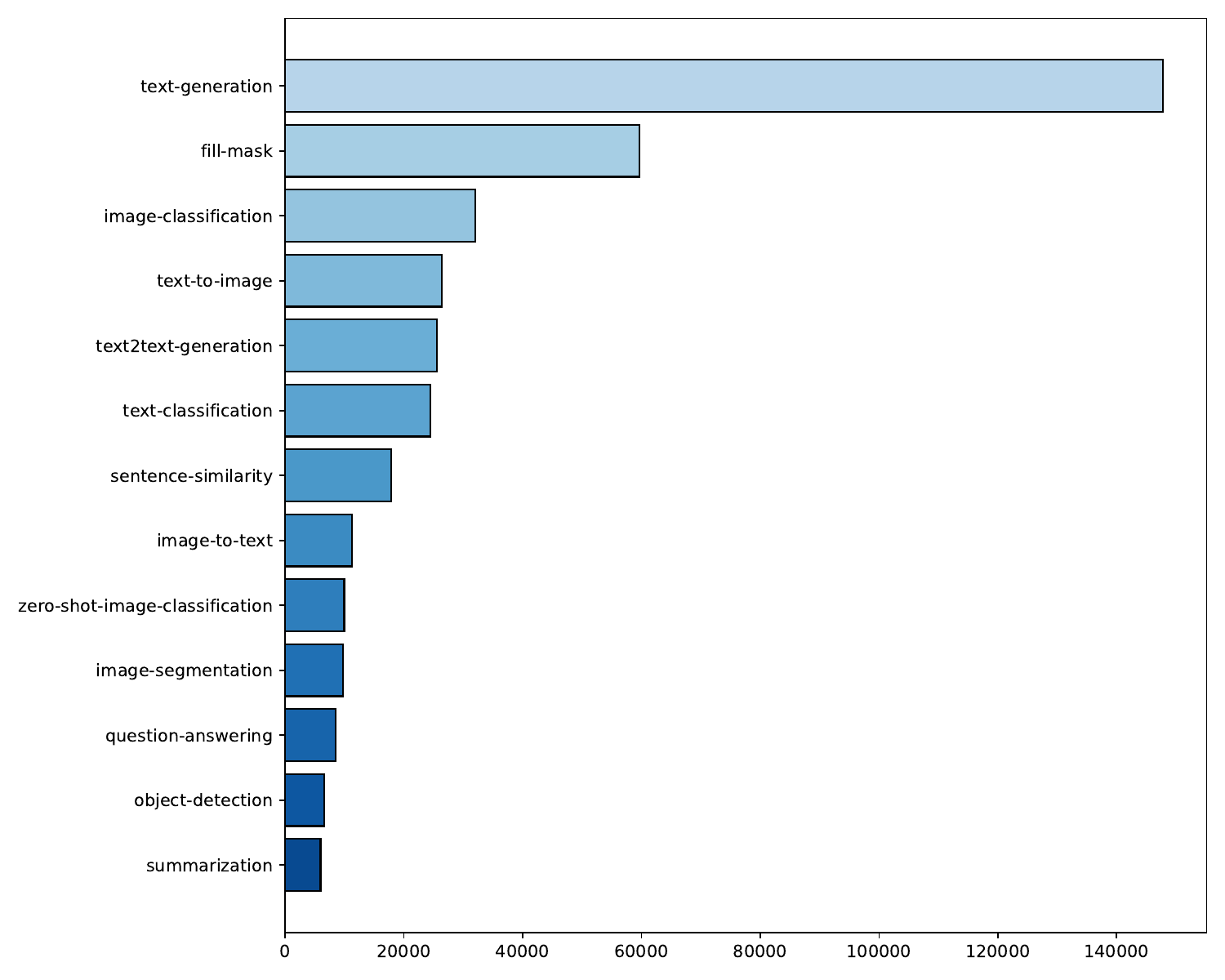}
    \caption{Num. of files for each tag in \tool}
    \label{fig:tags}
\end{figure}

This section presents an overview of \tool data by providing basic descriptive statistics. In particular, we aim to investigate two aspects, \ie the number of files of each indetified pipeline tag  and the usage of the PTMs in the \GH ecosystem.
Concerning the first aspect, we present the distribution of the number of files for each tag in \tool. As shown in Figure \ref{fig:tags}, the file distribution is unbalanced since some tags are more used compared to others. In particular, the most popular tag is \textit{text-generation}, followed by \textit{fill-mask} and \textit{image-classification}. This is quite expected since the most popular PTMs, like \textit{gpt-2} and \textit{RoBERTa} model, are tagged with text-generation and fill-mask, respectively. Nevertheless, we include also less popular tags to support the development of a wide range of applications based on PTMs, thus increasing the diversity of the dataset.

\begin{figure}[h!]
	\centering
	\begin{subfigure}[b]{0.79\linewidth}
		\centering
		\includegraphics[width=\linewidth]{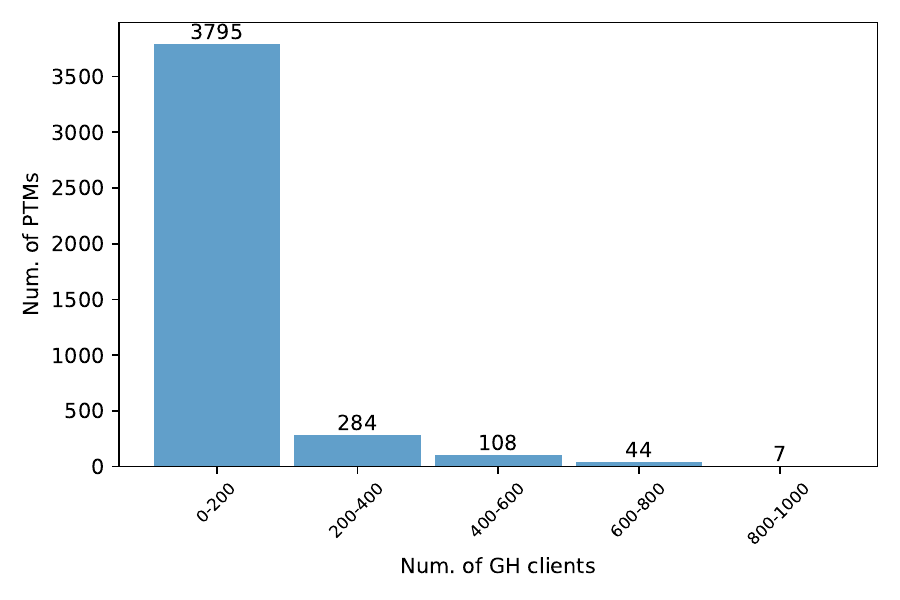}
		\caption{PTMs popularity }
		\label{fig:tail}
	\end{subfigure}
	\hfill\\
	\begin{subfigure}[b]{0.78\linewidth}
		\centering
		\includegraphics[width=\linewidth]{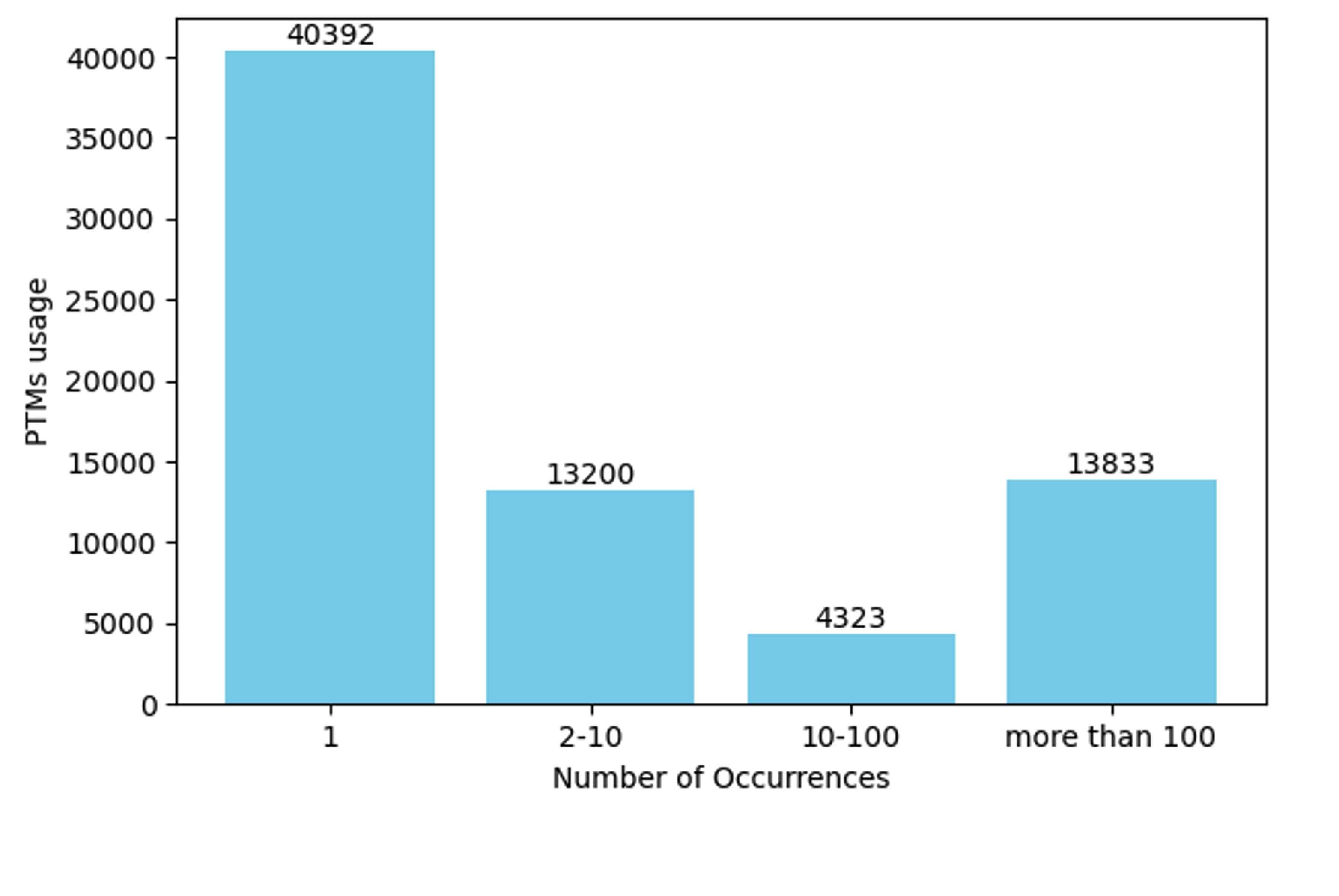}
		\caption{PTMs usage }
		\label{fig:repo-distribution}
	\end{subfigure}
	\caption{Distribution of PTMs in \tool}
	\label{fig:tail-repo}
\end{figure}

Concerning the popularity of the PTMs in the \GH ecosystem, Figure \ref{fig:tail-repo} reports the actual usage of the collected PTMs. In particular, we investigate the distribution of the number of projects for each PTM depicted in Figure \ref{fig:tail}. The result shows a skewed distribution that exibiths a \textit{long-tail} effect, with a small number of PTMs used in a large number of \GH projects. By carefully inspecting the data, we discovered that the most popular PTMs are \textit{gpt-2}, \textit{roberta-large}, and \textit{Salesforce/blip-image-captioning-large}, which are used in more than 800 projects. Thus, we can conclude that there are few PTMs that are widely used in the \GH ecosystem, while the majority of the PTMs are used in a small number of projects.
 
In addition, we investigated the how many PTMs \GH repositories include. Figure~\ref{fig:repo-distribution} shows that a considerable number of repositories (43.7\%) uses more than one PTMs, meaning that developers may combine different models to support one or more tasks. Interestigly, we spot repositories that use more than 100 models,  \eg hugging-downloader\footnote{\url{/https://github.com/isLinXu/hugging-downloader}}. By carefully investigate this, we discovered that those kinds of projects are model downloaders employed to store and test many PTMs at once.

\label{sec:analysis}


\section{Predict PTM usage patterns}
Building upon our novel dataset that establishes a crucial link between \HGF model cards and their corresponding usage in publicly available source code, this section explores an illustrative application of this resource. While not the central focus of this paper, this explanatory use case demonstrates that \tool can be used to support the prediction of code usage patterns. First, we filter out outliers and applied a clustering technique to indetify the most suitable code patterns. Then, we expolit  the Llama model to generate code snippets based on the identified patterns.

\subsection{Outlier filtering}
As reveal by our previous analysis, the collected data presents a long-tail effect, \ie a small subset of models is employed far more frequently than the majority. Thus, we filter out outliers from our analysis to those pre-trained models (PTMs) associated with more than 100 files. This criterion yielded a subset of 1,064 models, encompassing a total of 393,484 script files. Table \ref{tab:outlier_filter} summarizes the key statistics of the original and filtered dataset. 

\begin{table}[h]
    \centering
    \caption{Number of file for each PTM}
    \label{tab:outlier_filter}
    \begin{tabular}{|l|c|c|}
        \hline
        \textbf{Metric} & \textbf{Value (Original)}& \textbf{Value (Filtered)} \\
        \hline
        Mean & 63.99 & 369.82  \\
        Median & 6.00 & 272.50\\
        Standard deviation & 162.54 &  265.65 \\
        Skewness & 3.86 & 1.17 \\
        Kurtosis & 16.09 & 0.31\\
        \hline
    \end{tabular}

\end{table}

In particular, the process increase the number of supportin file for each PTM on average, from 63.99 to 369.82. In addition, skewness and kurtosis decreased to 1.17 and 0.31, respectively.  Overall, this filtering step contributes to a more balanced distribution of the data, which is crucial for generating accurate snippets. To further enhance the quality of the code snippets, we applied an additiona filter on the size of each snippet. On the one hand, excessively brief scripts may lack substantive content regarding a model’s usage. On the other hand, long scripts often contain non‐pertinent material \eg extensive documentation or commentar. Thus, we restrict our analysis to those scripts whose length, measured in lines of code, falls within a specified interquartile range (IQR) centered on the median~\cite{vinutha2018detection}, defined as follows:

\begin{equation}
    min\_length =max(mediana-0.25*IQR,1)
\end{equation}
\begin{equation}
    max\_length =mediana+0.25*IQR
\end{equation}

\subsection{Clustering Analysis}

After the selection of the most similar snippets, we grouped snippets that contain at least one match to the predefined keywords. However, selecting the optimal candidate requires accounting for the semantic content of the code. A similar strategy was employed by Zhong et al.\cite{zhong2009mapo}, who improve recommendation quality by clustering code snippets according to analogous API usage patterns.

Thus, we applied the K-means algorithm from the scikit-learn library\cite{scikit-learn} to numerical representations of the snippets. Since the snippets are initially plain text, we employed the all-MiniLM-L6-v2\footnote{\url{https://huggingface.co/sentence-transformers/all-MiniLM-L6-v2}} embedding model provided by Hugging Face via the sentence\_transformers package\footnote{\url{https://sbert.net/}}. This model based on a lightweight Transformer variant of BERT—converts each snippet into a 384-dimensional vector through tokenization, attention computation, and mean-pooling across token embeddings. Although sentence\_transformers models are primarily optimized for natural language, they nevertheless capture substantial syntactic and semantic information in code fragments, particularly when natural-language comments elucidate code structures and functions. Moreover, their relatively modest inference time and memory footprint present a favorable trade-off for processing hundreds of thousands of files, in contrast to more heavyweight alternatives such as CodeBERT \cite{feng2020codebertpretrainedmodelprogramming}.

The resulting embeddings are then clustered via K-means, with the number of clusters
k chosen empirically as the minimum of three and the total number of snippet candidates; this heuristic ensures dynamic adaptation to the available snippet set. Upon convergence, the snippet whose embedding lies closest to each cluster centroid is selected as the representative exemplar. By virtue of its proximity to the centroid, this snippet best encapsulates the shared semantic characteristics of its cluster members.

\subsection{Code pattern prediction}

To perform code generation, we exploit LLM for generating code usage example. Noteworthy, the dataset and the whole process can be adopted in combination with more traditional techniques, \ie collaborative filtering \cite{Fowkes:2016:PPA:2950290.2950319,Nguyen:2019:FRS:3339505.3339636} or mining usage pattenrs \cite{6624045,10.1007/s10664-021-10000-w}. In the scope of this paper, we employ Llama 3 model \cite{grattafiori2024llama3herdmodels} with a zero-shot prompting paradigm, wherein the model receives no explicit exemplars of the desired output within the input string and must rely solely on its inherent generalization capabilities. As detailed in Listing \ref{llm_prompt}, the prompt further assigns the role of “Python assistant” to the model, thereby optimizing contextual relevance. In additoin, the prompt includes constraints to avoid the generation of natural-language content. 


Concerning the hyperparamers, we set a temperature parameter of 0.2, the top\_p parameter equal to 1.0, meaning that token selection considers the full probability distribution produced by the model. By avoiding the truncation inherent in lower top\_p thresholds—which limit selection to tokens within a cumulative-probability cutoff—this configuration fully leverages the model's capabilities without prematurely excluding less probable yet potentially valid alternatives. Given that the goal is precise identification of code patterns rather than creative generation, this low temperature reduces response variance and mitigates the risk of spurious or hallucinatory outputs.

\begin{lstlisting} [language=json, caption={The zero-shot prompt used to generate model card usage patterns}, label={llm_prompt}]
    system_prompt = (
    "You are an AI assistant specialized in Python.
    I will provide you with a series of code snippets extracted from various Python files.
    Note that these snippets might not form a coherent or complete code module when combined together.
    Your task is to analyze these snippets and extract recurring code patterns, such as function definitions (using 'def'), common imports, and other typical structures found in Python code.
    Focus solely on identifying and returning the common patterns in the context of the model {snippet_model}.
    Return ONLY the code snippets.
    DO NOT include only the common imports.
    Do NOT include any explanations, descriptions, or metadata.
    Do NOT generate any summaries, bullet points, markdown formatting, or additional text.
    If you are unable to find any relevant patterns, please return an empty string."
    )    
    \end{lstlisting}

Listing \ref{llm_example2} presents an illustrative code excerpt generated by the LLM for the model kakaobrain/karlo-v1-alpha-image-variations, a framework typically employed to synthesize images from natural-language prompts\footnote{\url{https://huggingface.co/kakaobrain/karlo-v1-alpha-image-variations}}. 
In particular, the generated code support the data loading, suggesting a standard procedure for ingesting such datasets into memory (lines 1-4).
Then, a preprocessing pipeline (lines 6–10) is established, by including a sequence of image transformations is applied—namely, resizing and normalization—ensuring that inputs conform to the model’s expected dimensionality and value range. Finally, the snippet outlines the construction of a convolutional encoder–decoder architecture leveraging the Keras API\footnote{\url{https://keras.io/}} as shown in Lines 12-24. 
While this example may diverge in detail from the official Karlo model card, it provides a relevant usage patterns that can help developers in integrating or extending the within their active projects.


\begin{lstlisting}[language=json, caption={Explanatory output for kakaobrain/karlo-v1-alpha-image-variations model }, label={llm_example2}]
def load_data():
    # Load data from kakaobrain/karlo-v1-alpha-image-variations
    data = pd.read_csv('data/train.csv')
    return data

def preprocess_image(image):
    # Resize and normalize image
    image = tf.image.resize(image, (224, 224))
    image = tf.image.per_image_standardization(image)
    return image

def create_model(input_shape):
    # Create Keras model
    model = tf.keras.models.Sequential([
        tf.keras.layers.Conv2D(32, (3, 3), activation='relu', input_shape=input_shape),
        tf.keras.layers.MaxPooling2D((2, 2)),
        tf.keras.layers.Conv2D(64, (3, 3), activation='relu'),
        tf.keras.layers.MaxPooling2D((2, 2)),
        tf.keras.layers.Conv2D(128, (3, 3), activation='relu'),
        tf.keras.layers.Flatten(),
        tf.keras.layers.Dense(128, activation='relu'),
        tf.keras.layers.Dense(10, activation='softmax')
    ])
    return model

def train_model(model, data):
    # Train model
    model.compile(optimizer='adam', loss='sparse_categorical_crossentropy', metrics=['accuracy'])
    model.fit(data, epochs=10)
    return model

def evaluate_model(model, data):
    # Evaluate model
    loss, accuracy = model.evaluate(data)
    return loss, accuracy

def save_model(model):
    # Save model
    model.save('model.h5')
      
\end{lstlisting}

\label{sec:usage}

\section{Discussion}
This section forseen possible usages of the \tool dataset. The primary usage of \tool is generating enhanced version of model card, \ie providing relevant snippets of code. In Section~\ref{sec:usage}, we demonstrated that curated model card can be used to automatically generate enhanced model card with code usage leveraging the provided source code from \GH. However, preprocessing and filtering steps are still needed to provide a high-quality model card as the raw data exhibited a significant long-tail effect.
Subsequentely, there is the need to filter out outliers and analyze the frequency of model usage across a large corpus of source code to effectively generate a useful model card. In this respect, the evaluation of our proposed approach for automated model usage recommendation relies on the insights derived from this dataset although an in-depth analysis is needed to confirm our findings. To understand the performances of the envisioned approach in Section~\ref{sec:usage}, a direct comparison between the textual information in the cards and the patterns observed in the linked source code is essential. Overall, our curated dataset can be effectively used for the three abovementioned tasks. 

Besides the model card generation, we belive that \tool can be used to  
the \tool dataset can be used to investigate how PTMs are used in real-world projects on \GH, thus fostering application of different automated techiques to support developers that want to implement, fine-tune, or test their wown models. In particular, \tool dataset can be used to provide different types of recommendations, spanning from API function calls \cite{Fowkes:2016:PPA:2950290.2950319,Nguyen:2019:FRS:3339505.3339636} to code usage pattern retrieval \cite{6624045,10.1007/s10664-021-10000-w}, providing a specific support to developers interested in integrating PTMS in their software projects. In addition, our dataset can be used to analyze how PTMs are used in real-world projects. In particular, the retrieved code can be used to assess the overall quality of the code, \eg code smells \cite{10.1145/3180155.3182532}, techinical debt in AI-based system \cite{RECUPITO2024112151}, or anti-patterns \cite{khomh_exploratory_2012}. In addition, the dataset can be used to investigate the impact of PTMs on the \GH community, \eg how the adoption of PTMs affects the project's popularity, its quality, and discoverability. 

\label{sec:opportunities} 

\section{Threats to validity}
This section discusses threats that may hamper the quality of the collected data. Concerning the \textit{internal validity}, the main issue is related to the dump used for collecting \tool, \ie the HF dump may not contain recently released PTMs. To mitigate this issue, we focus on popular and well-established models by selecting PTMs. In addition, we applied different quality filter to reduce the number of the outliers. Concerning the generation of usage patters scenario presented in Section~\ref{sec:usage}, we acknowledge that the clustering technique used to group similar snippets may not be the most suitable for all scenarios. To mitigate this, we plan to investigate more advanced clustering techniques in future work. We also acknowledge that employed LLM,\ie Llama, may not be the best choice for all tasks, thus generating erroneous or irrelevant code snippets. To handle this, we first filter out outliers and used a curated prompt.
\textit{External validity} is related to missing data that may not be collected during the GH mapping phase. To handle this, we rely on the GitHub query language to apply quality filters. In addition, we manually select six categories to include less popular PTMs in the final dataset. Another threat is related to the possible usage of the dataset, \ie the PTM snippets may be not enough to train a traditional model. To mitigate this, we employ a clustering technique to group similar snippets and increase the amount of data. Another threat is related to the generalizability of the results, \ie the dataset may not provide enough examples to support the generation of generic model cards. While we acknowledge that this can be mitigated by enlarging the dataset with additional PTMs, this goes beyond the scope of this paper as we focus on the collection and curation of the retrieved data from HF and \GH. 
\label{sec:threats}

\section{Related works}
In this section, we present existing empirical studies on PTMs in Section~\ref{sec:empirical-studies} while Section~\ref{sec:code-usage-pattern} reviews existing woks on recommending code usage pattern.
\subsection{Emprical Studies on PTMs}\label{sec:empirical-studies}

Castano \etal \cite{castano_analyzing_2023} investigate the carbon footprints of 1,417 different models hosted on the platform. The measured emission correlated with different factors such as model size, dataset size, and application domains. The same authors \cite{castano_analyzing_2023} also provide an empirical investigation on the evolution of PTMs in terms of maintenance, popularity, and usage. Gong \etal \cite{gong_what_2023} conducted a comprehensive study of the PTMs reuse stored in six different model repositories, including \HGF. After data cleaning and labeling steps, the authors propose a code contract composed of  pre- and post-conditions for re-usage in software development, \eg input data, intended usage, and performance. Montes \etal \cite{montes_discrepancies_2022} highlights discrepancies in the documentation of 36 PTMs that support image classification across four different model repositories, \ie TensorFlow Model Garden, ONNX Model Zoo, Torchvision Models, and Keras Applications, highlighting the need for standardized documentation. Pepe \etal \cite{10.1145/3643916.3644412} conducted a large-scale study on 159,132 models stored on HF by focusing on the documentation, licenses, and fairness aspects. Overall, only a few PTMs provide permissive licenses and mention potential bias in the documentation. Gao \etal \cite{10.1145/3674805.3686679} investigate ethical concerns in HF models leveraging the API and KeyBERT model, ending up with a dedicated taxonomy. 
 In our prior work \etal \cite{10.1145/3661167.3661215}, we investigate to what extent traditional ML models, namely Naive Bayesian and SVC, can classify PTM given their model card. 
 Compared with the abovementioned works, \tool provides a direct mapping between PTMs and \GH projects that use them in practice.

 \subsection{Recommending code usage patterns}\label{sec:code-usage-pattern}

FACER \cite{abid_facer_2021} retrieves API usage for opportunistic reuse built on top of a code fact repository, including methods’ textual body, call graphs, and API
usages. The approach combines custering technique based on Lucene and frequent pattern mining strategy to retrieve similar API compared to the developer's context.

 \textit{api2vec}~\cite{nguyen_exploring_2017} is an approach based on Word2Vec model to suggest relevant API patterns and usage. First, the tool mines the pairs of API elements that share the same usage relations among them. Afterward, the mined content has been used to support code translation task between Java and C\# using a characteristic of the API2Vec embeddings.

 Multi-HyLSTM~\cite{xiao_specializing_2023} is an automated approach that exploits a modified version Short Long Term Memory (LSTM) neural network to support multi-path API prediction. The system is equipped with global dependence-enhancing learning module to accurately capture the program dependencies for an API calls. 





 To enhance the prediction of LLM like GPT-3 and Codex, Jain \etal propose Jigsaw ~\cite{jain_jigsaw_2022}, a tool based on program synthesis architecture to augment the input of LLM in code generation task. In particular, it contextualizes the input to
 the black-box language model using heuristic techniques. Afterward, a post-processing module speed up the combinatorial search
 space of API functions and their arguments leveraging API usage example provided by humans. Specifically trained on a Pandas dataset, Jigsaw improves the quality of the generated code of GPT-3 and Codex.

 Compared to those work, \tool can be seen as a curated source of knowledge to enable tailored recommendations for PTMs and their usage patterns. In particular, we provide a dataset of PTMs and their related code usage patterns, which can be used to train and evaluate different recommendation systems.
%

\label{sec:related}

\section{Conclusion}
Motivated by increasing usage of pre-trained models (PTMs) for prominent software engineering tasks, we propose \tool, a curated dataset of PTMs exploited in the \GH ecosystem. First, we identified the most downloaded PTMs from the Hugging Face model repository dump and filtered out PTMs with missing content. Afterward, we leverage the PyGithub library to collect the actual usage of the identified PTMs in \GH projects. We ended up with \numWithCode different models and \numFiles Python files. We also present a statistical analysis of the dataset, highlighting the most popular PTMs and the most common tasks for which they are used. In addition, we demonstrate a concrete application of \tool, focusing on extracting representative code usage patterns for specific PTMs by combining clustering techniques and Large Language Models (LLMs). We also discuss the research opportunities enabled by \tool and the implications of our findings for the software engineering community. In particular, we believe that \tool can be used to support a plethora of tasks, ranging from API recommendation to generating enhanced model cards, contributing to the development of next-generation solutions based on top of PTMs.


For future work, we plan to extend the dataset by leveraging new dumps released by the HF community project or Hugging Face dedicated API. In addition, we can mine source code from additional open-source repositories, \eg Software Heritage or GitLab, or model repositories, \eg TensorFlow Model Garden or ONNX Model Zoo, and their usage in the GitHub platform. Concering the prediction of PTM usage patterns, we plan to investigate the use of more advanced clustering techniques, such as deep learning-based approaches, and experiment with different LLMs to generate more accurate and relevant code usage patterns. Finally, we plan to conduct a user study to evaluate the effectiveness of the proposed approach in real-world PTM-based project.
\label{sec:conclusion}


\bibliographystyle{IEEEtran}
\bibliography{main}

\end{document}